\documentstyle[epsfig]{aipproc}

\def\beq{\begin{equation}}
\def\eeq{\end{equation}}
\def\beqar{\begin{eqnarray}}
\def\eeqar{\end{eqnarray}}
\def\barr#1{\begin{array}{#1}}
\def\earr{\end{array}}
\def\bfi{\begin{figure}}
\def\efi{\end{figure}}
\def\btab{\begin{table}}
\def\etab{\end{table}}
\def\bce{\begin{center}}
\def\ece{\end{center}}

\def\text{\textstyle}

\def\arraystretch{1.4}

\def\al{\alpha}

\def\De{\Delta}

\def\reffi#1{\mbox{Fig.~\ref{#1}}}

\def\citere#1{\mbox{Ref.~\cite{#1}}}
\def\citeres#1{\mbox{Refs.~\cite{#1}}}

\def\mathswitchr#1{\relax\ifmmode{\mathrm{#1}}\else$\mathrm{#1}$\fi}

\newcommand{\PW}{\mathswitchr W}
\newcommand{\PZ}{\mathswitchr Z}
\newcommand{\PA}{\mathswitchr A}

\newcommand{\PH}{\mathswitchr H}

\newcommand{\Ph}{\mathswitchr h}

\newcommand{\Pb}{\mathswitchr b}

\newcommand{\Pt}{\mathswitchr t}

\def\mathswitch#1{\relax\ifmmode#1\else$#1$\fi}

\newcommand{\MW}{\mathswitch {M_\PW}}

\newcommand{\Mt}{\mathswitch {m_\Pt}}

\newcommand{\mh}{\mathswitch {m_\Ph}}
\newcommand{\mH}{\mathswitch {m_\PH}}
\newcommand{\MA}{\mathswitch {M_\PA}}

\newcommand{\GF}{\mathswitch {G_\mu}}

\def\tb{\tan\beta}

\newcommand{\Xt}{X_{\Pt}}

\newcommand{\aeff}{\al_{\mathrm{eff}}}

\newcommand{\mt}{\Mt}

\newcommand{\mgl}{m_{\tilde{\mathrm{g}}}}

\newcommand{\Stop}{\tilde{\Pt}}

\newcommand{\Sbot}{\tilde{\Pb}}

\newcommand{\tsf}{\theta\kern-.20em_{\tilde{f}}}
\newcommand{\tsfp}{\theta\kern-.20em_{\tilde{f}\prime}}
\newcommand{\tsq}{\theta\kern-.15em_{\tilde{q}}}

\newcommand{\msusy}{M_{\mathrm{SUSY}}}
\newcommand{\mslep}{M_{\mathrm{\tilde l}}}

\newcommand{\lsim}
{\;\raisebox{-.3em}{$\stackrel{\displaystyle <}{\sim}$}\;}

\newcommand{\alps}{\alpha_{\mathrm s}}

\newcommand{\feh}{{\em FeynHiggs}}
\newcommand{\subh}{{\em subhpole}}

\newcommand{\cp}{{\cal CP}}

\newcommand{\VL}{\left( \begin{array}{c}}
\newcommand{\VR}{\end{array} \right)}
\newcommand{\ML}{\left( \begin{array}{cc}}
\newcommand{\MLd}{\left( \begin{array}{ccc}}
\newcommand{\MLv}{\left( \begin{array}{cccc}}
\newcommand{\MR}{\end{array} \right)}

\newcommand{\BC}{\begin{center}}
\newcommand{\EC}{\end{center}}
\newcommand{\BE}{\begin{equation}}
\newcommand{\EE}{\end{equation}}
\newcommand{\BEA}{\begin{eqnarray}}
\newcommand{\BEAnn}{\begin{eqnarray*}}
\newcommand{\EEA}{\end{eqnarray}}
\newcommand{\EEAnn}{\end{eqnarray*}}

\newcommand{\id}{{\rm 1\kern-.12em
\rule{0.3pt}{1.5ex}\raisebox{0.0ex}{\rule{0.1em}{0.3pt}}}}

\def\draftdate{\relax}
\def\mda{\relax}
\def\mua{\relax}
\def\mla{\relax}
\def\draft{
\def\thtystars{******************************}
\def\sixtystars{\thtystars\thtystars}
\typeout{}
\typeout{\sixtystars**}
\typeout{* Draft mode!
         For final version remove \protect\draft\space in source file
*}
\typeout{\sixtystars**}
\typeout{}
\def\draftdate{\today}
\def\mua{\marginpar[\boldmath\hfil$\uparrow$]%
                   {\boldmath$\uparrow$\hfil}%
                    \typeout{marginpar: $\uparrow$}\ignorespaces}
\def\mda{\marginpar[\boldmath\hfil$\downarrow$]%
                   {\boldmath$\downarrow$\hfil}%
                    \typeout{marginpar: $\downarrow$}\ignorespaces}
\def\mla{\marginpar[\boldmath\hfil$\rightarrow$]%
                   {\boldmath$\leftarrow $\hfil}%
                    \typeout{marginpar:
$\leftrightarrow$}\ignorespaces}
\def\Mua{\marginpar[\boldmath\hfil$\Uparrow$]%
                   {\boldmath$\Uparrow$\hfil}%
                    \typeout{marginpar: $\Uparrow$}\ignorespaces}
\def\Mda{\marginpar[\boldmath\hfil$\Downarrow$]%
                   {\boldmath$\Downarrow$\hfil}%
                    \typeout{marginpar: $\Downarrow$}\ignorespaces}
\def\Mla{\marginpar[\boldmath\hfil$\Rightarrow$]%
                   {\boldmath$\Leftarrow $\hfil}%
                    \typeout{marginpar:
$\Leftrightarrow$}\ignorespaces}
\overfullrule 5pt
\oddsidemargin -15mm
\marginparwidth 29mm
}

\begin{document}

\thispagestyle{empty}
\setcounter{page}{0}
\def\thefootnote{\fnsymbol{footnote}}

\begin{flushright}
BNL--HET--01/5\\
CERN--TH/2001-037\\
hep-ph/0102117 \\
%\date{\today}
\end{flushright}

\vspace{1cm}

\begin{center}

{\large\sc {\bf Higgs Production and Decay in the MSSM:}}

\vspace*{0.4cm} 

{\large\sc {\bf Status and Perspectives}}
%\footnote{talk given by authorXY at}

\vspace{1cm}

{\sc S. Heinemeyer$^{1\,}$%
\footnote{
email: Sven.Heinemeyer@bnl.gov
}%
~and G. Weiglein$^{2\,}$%
\footnote{
email: Georg.Weiglein@cern.ch
}%
}

\vspace*{1cm}

$^1$ HET, Brookhaven Natl.\ Lab., Upton, New York 11973, USA

\vspace*{0.4cm}

$^2$ CERN, TH Division, CH-1211 Geneva 23, Switzerland

\end{center}

\vspace*{1cm}

\begin{abstract}
The theoretical predictions in the MSSM for Higgs-boson production at a
future $e^+e^-$ Linear Collider and Higgs-boson decay processes are 
discussed focusing in particular on
recent diagrammatic two-loop results in the MSSM Higgs sector.
The present status of the theoretical predictions is briefly summarized,
and it is emphasized that considerable improvements will be necessary in
order to match the high experimental accuracy achievable at a future
Linear Collider.
\end{abstract}

\def\thefootnote{\arabic{footnote}}
\setcounter{footnote}{0}

\newpage

%%%%%%%%%%%%%%%%%%%%%%%%%%%%%%%%%%%%%%%%%%%%%%%%%%%%%%%%%%%%%%
%%%%%%%%%%%%%%%%%%%%%%%%%%%%%%%%%%%%%%%%%%%%%%%%%%%%%%%%%%%%%%

\title{Higgs production and decay in the MSSM: status and perspectives}

\author{Sven Heinemeyer$^*$ and Georg Weiglein$^{\dagger}$}
\address{$^*$HET, Brookhaven Natl.\ Lab., Upton, New York 11973, USA\\
$^{\dagger}$CERN, TH Division, CH-1211 Geneva 23, Switzerland}

%\lefthead{LEFT head}
%\righthead{RIGHT head}
\maketitle

\begin{abstract}
The theoretical predictions in the MSSM for Higgs-boson production at a
future $e^+e^-$ Linear Collider and Higgs-boson decay processes are 
discussed focusing in particular on
recent diagrammatic two-loop results in the MSSM Higgs sector.
The present status of the theoretical predictions is briefly summarized,
and it is emphasized that considerable improvements will be necessary in
order to match the high experimental accuracy achievable at a future
Linear Collider.
\end{abstract}

\section{Introduction}

%The existence of a light neutral Higgs boson is an important prediction
%of Supersymmetry (SUSY), which can be tested at the next generation of
%colliders. 
Exploring the Higgs sector of the electroweak theory with
high precision will be one of the main goals of a future $e^+e^-$ Linear
Collider (LC). It will allow a highly sensitive test of the investigated
theory and will 
thus provide a way for distinguishing between different models.

A firm prediction of the Minimal Supersymmetric Standard Model (MSSM)
is the existence of a light neutral Higgs boson. The Higgs sector of the MSSM
consists of two doublets and besides the gauge couplings is described by
two parameters, conventionally chosen as the ratio of the two vacuum
expectation values, $\tb = v_2/v_1$, and the mass of the $\cp$-odd Higgs
boson, $\MA$. The masses of the other Higgs bosons and the mixing angle
in the neutral Higgs sector are predicted in terms of these parameters,
giving rise to the well-known result that the mass of the 
lightest $\cp$-even Higgs boson, $\mh$, in the MSSM has to be smaller than the
Z-boson mass at lowest order. This bound, however, receives large
radiative corrections~\cite{mhiggsrad}, in particular from the Yukawa 
sector of the theory, which shift the bound to $\mh \lsim 135$~GeV including 
two-loop corrections~\cite{mhiggslong}.

The prospective experimental accuracies at a future Linear Collider in 
the energy range 500~GeV -- 1~TeV will make it necessary to reduce the
theoretical uncertainties of the predictions for the Higgs production
cross sections and branching ratios below the level of 1\%. This goal
will be very difficult to achieve, owing to the fact that there exist
several different sources for sizable higher-order corrections and for
large theoretical uncertainties. 

As mentioned above, large Yukawa corrections affect the predictions for
the Higgs-boson masses and the mixing angle $\al$. The leading terms at one-
and two-loop order are of ${\cal O}(\GF \mt^4/\MW^2)$ and ${\cal O}(\GF
\alps \mt^4/\MW^2)$, respectively. Besides these corrections, which
originate from the $\Pt$--$\Stop$ sector of the MSSM, for large values
of $\tb$ and the Higgs mixing parameter $\mu$ also large corrections in
the $\Pb$--$\Sbot$ sector are possible, which in some regions of the
parameter space can even invalidate a perturbative treatment. 
Corrections from the scalar quarks of the third generation can also 
be a source for large loop-induced effects in the Higgs sector
connected to complex phases.
Concerning theoretical uncertainties related to large loop corrections,
in the energy range of a LC genuine vertex and box corrections in general 
give rise to much bigger effects than at LEP energies, where often the 
bulk of the weak and QCD corrections originates from universal
propagator-type contributions.

In the parameter regions where the widths of the MSSM Higgs bosons are 
large, an accurate treatment of the Higgs-boson production furthermore
requires to take into account off-shell effects, i.e.\ to study the full
$2 \to 4$ process including non-resonant contributions. 

In addition to the above-mentioned effects, a different source of
theoretical uncertainties is related to the experimental errors of the
input parameters. Owing to the large corrections from the $\Pt$--$\Stop$
sector, in particular an accurate measurement of the top-quark mass is
crucial for precise theoretical predictions in the Higgs sector of the
MSSM.

%%%%%%%%%%%%%%%%%%%%%%%%%%%%%%%%%%%%%%%%%%%%%%%%%%%%%%%%%%%%%%%%%%%%%
%%%%%%%%%%%%%%%%%%%%%%%%%%%%%%%%%%%%%%%%%%%%%%%%%%%%%%%%%%%%%%%%%%%%%

\section{The $\cp$-even Higgs-boson masses and $\aeff$}

The prediction for the lightest $\cp$-even Higgs-boson mass in the MSSM
has recently been improved by the inclusion of non-logarithmic 
genuine two-loop contributions obtained via an explicit
Feynman-diagrammatic (FD) calculation~\cite{mhiggslong,mhiggsFD}. These
corrections gave rise to a numerically sizable shift compared to the
results previously obtained via a renormalization-group-improved
one-loop Effective Potential approach (EPA)~\cite{mhiggsRGa,mhiggsRGb}.
Recently further sub-leading two-loop electroweak contributions have been
obtained~\cite{mhiggsyuk}.

The current theoretical uncertainty in $\mh$ from unknown higher-order 
contributions can conservatively be estimated as 
$\De\mh^{\mathrm{theo}} \approx \pm 3$~GeV.
This uncertainty is smaller than the one induced from the present
experimental error on $\mt$, as a change of $\De\mt = \pm 5$~GeV
leads to a shift in $\mh$ of 
$\De\mh^{\mathrm{theo}} \approx \pm 5$~GeV~\cite{tbexcl}. 
These uncertainties have to be compared with the prospective accuracies
at a future LC of $\De\mh^{\mathrm{exp}} = 0.05$~GeV and
$\De\mt^{\mathrm{exp}} = 0.1$--0.2~GeV. 

The theoretical uncertainties in the Higgs propagator corrections affect
the predictions for the Higgs production and decay processes via their 
effects on $\mh$, $\mH$, and on the effective mixing angle $\aeff$ in
which the bulk of the Higgs propagator corrections to the Higgs
couplings can be absorbed.

The two-loop Higgs propagator corrections evaluated in 
\citeres{mhiggslong,mhiggsFD} have recently been incorporated into the
diagrammatic one-loop result for the QED and QCD contributions~\cite{hff1l}
to the decay processes $h \to f \bar f$~\cite{hff} and into the complete
diagrammatic one-loop result~\cite{eehZA1l} for the production processes 
$e^+e^- \to hZ$ and $e^+e^- \to hA$~\cite{eeZhA}.

%%%%%%%%%%%%%%%%%%%%%%%%%%%%%%%%%%%%%%%%%%%%%%%%%%%%%%%%%%%%%%%%%%%%%
%%%%%%%%%%%%%%%%%%%%%%%%%%%%%%%%%%%%%%%%%%%%%%%%%%%%%%%%%%%%%%%%%%%%%

\section{Higgs decays into SM fermions}

\begin{figure}[ht]
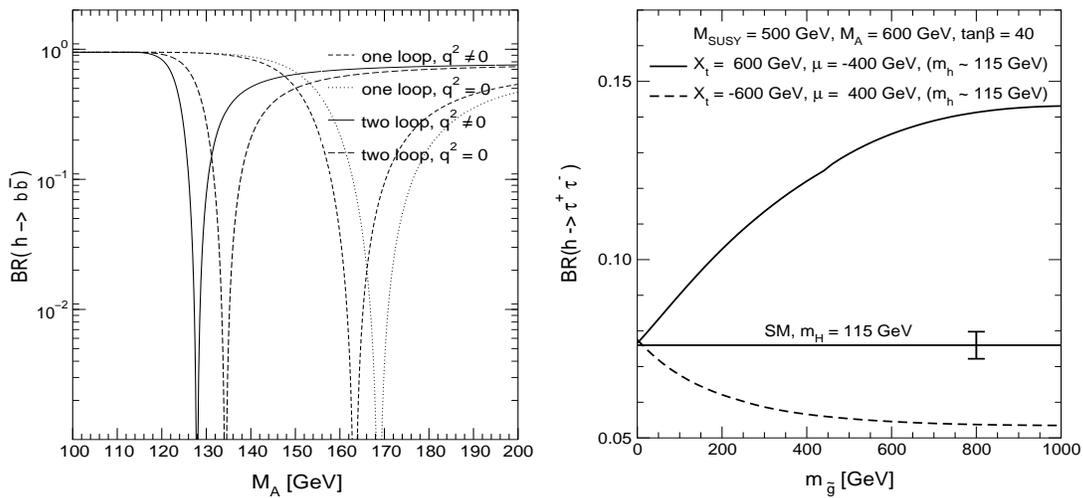

\begin{center}
\mbox{
\epsfig{figure=brh2lA5.bw.eps,width=7cm,height=6.5cm}}
\mbox{
\epsfig{figure=hff03.bw.eps,width=7cm,height=6.5cm}}
\end{center}
\caption[]{
Effects of large SUSY contributions on BR($h \to b \bar b$) (left plot)
and BR($h \to \tau^+ \tau^-$) (right plot). The error bar at the SM
prediction in the right plot indicates the prospective experimental 
accuracy at a future LC.
\label{fig:btaubr}
}
\end{figure}

The predictions for the decays $h \to f \bar f$ in the MSSM are affected
by two main sources of large corrections which can give rise to large
deviations of the MSSM predictions compared to the SM case. The Higgs
propagator corrections affect the couplings to fermions in particular 
via large $h$--$H$ mixing effects. As a consequence, the effective 
$h f \bar f$ coupling can be heavily suppressed in certain regions of
the parameter space~\cite{hff,patho1,patho2}. This is shown in the left 
plot of \reffi{fig:btaubr}, where BR($h \to b \bar b$) is given as a 
function of $\MA$ for $\msusy = \Xt = 500$~GeV and $\mu = - 1$~TeV 
($\msusy$ is the
squark mass scale and $\Xt$ is the off-diagonal entry in the $\Stop$ mixing 
matrix, see \citere{mhiggslong}). The plot shows that for a given value
of $\MA$ the theoretical prediction for BR($h \to b \bar b$) can change
drastically if a part of the corrections is neglected. This is illustrated
here by neglecting the two-loop contributions or the momentum dependence 
of the Higgs propagator corrections (the latter corresponds to the $\aeff$
approximation). As a consequence, in these regions of parameter space
an accurate theoretical prediction is very difficult to achieve.

%%%%%%%%%%%%%%%%%%%%% FIGURE %%%%%%%%%%%%%%%%%%%%%%%%%%%%%%%%%%%%%%%%%%
%%%%%%%%%%%%%%%%%%%%% FIGURE %%%%%%%%%%%%%%%%%%%%%%%%%%%%%%%%%%%%%%%%%%

While the Higgs propagator corrections affect the Higgs couplings to all 
up-type fermions and to all down-type fermions in a universal way, these
couplings can be shifted relative to each other by large
gluino and higgsino loop corrections. Corrections of this kind
can occur for large values of
$\tan\beta$ and/or $\mu$. They affect the tree-level relation between the 
fermion masses (in particular $m_b$ and $m_{\tau}$) and the Yukawa
couplings~\cite{deltab}. A deviation in the ratio of the $h b \bar b$
and the $h \tau^+ \tau^-$ couplings from the SM value caused by large 
gluino corrections to $\Gamma(h \to b \bar b)$ can give rise to 
a sizable shift in BR($h \to \tau^+ \tau^-$). A precise measurement
of BR($h \to \tau^+ \tau^-$) at a future LC will thus provide a high 
sensitivity for a distinction between the SM and the MSSM even for
relatively large values of $\MA$, where otherwise the Higgs sector
behaves mainly SM-like. This is illustrated in the right plot of
\reffi{fig:btaubr}, where BR($h \to \tau^+ \tau^-$) in the MSSM is shown 
as a function of the gluino mass, $\mgl$, in comparison with the
SM prediction and the prospective experimental accuracy at a future LC
of about 5\%.

%%%%%%%%%%%%%%%%%%%%%%%%%%%%%%%%%%%%%%%%%%%%%%%%%%%%%%%%%%%%%%%%%%%%%
%%%%%%%%%%%%%%%%%%%%%%%%%%%%%%%%%%%%%%%%%%%%%%%%%%%%%%%%%%%%%%%%%%%%%

\section{Higgs production in Higgs-strahlung and associated production}

\begin{figure}[ht]
\begin{center}
\mbox{
\epsfig{file=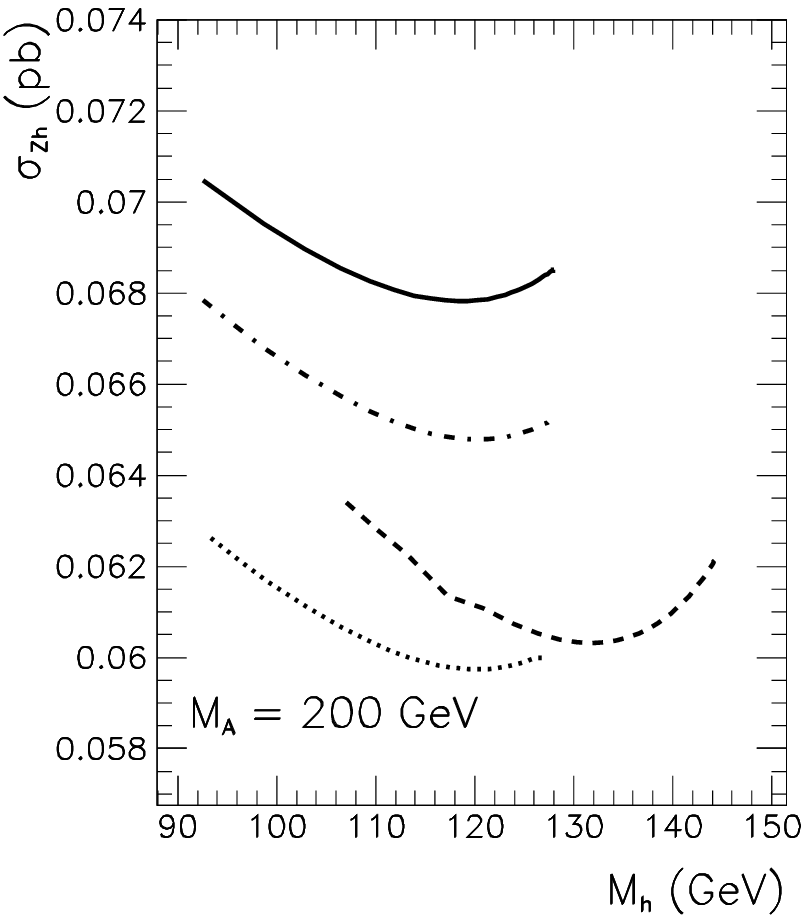,width=7.5cm,height=6.5cm}}
\hspace{-2em}
\mbox{
\epsfig{file=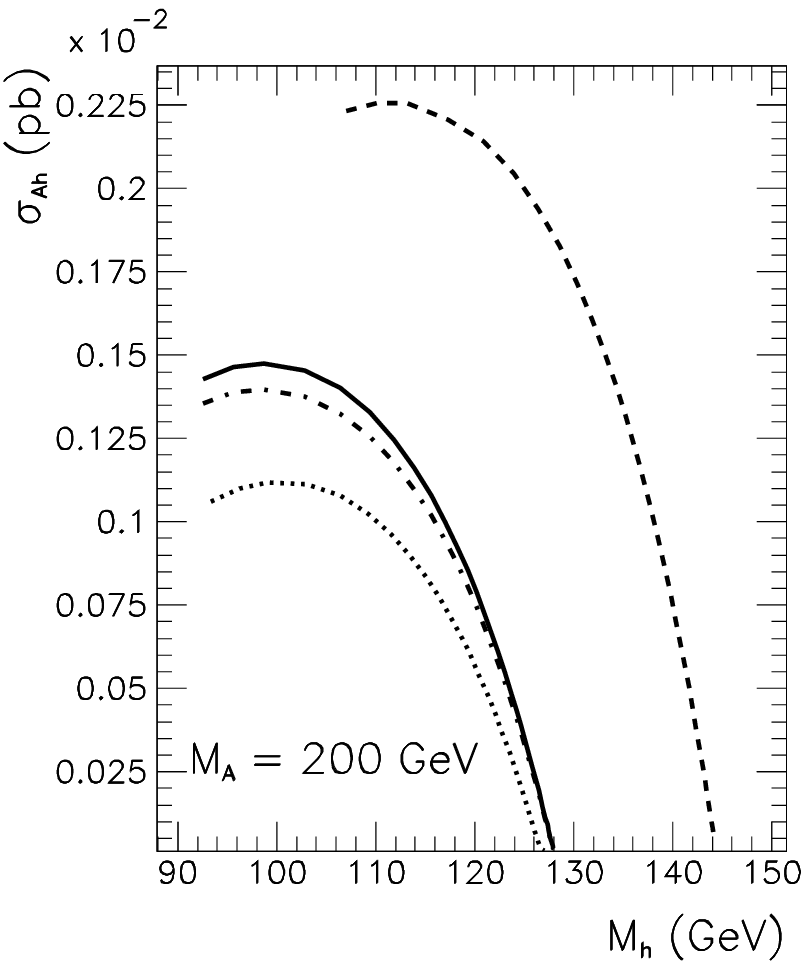,width=7.5cm,height=6.5cm}}
\end{center}
\caption[]{
$\sigma_{\PZ\Ph}$ and $\sigma_{\PA\Ph}$ as a function of $\mh$ at $\sqrt{s} =
500$~GeV for maximal mixing in the scalar top sector. The solid
line represents the FD result consisting of the complete
one-loop and the two-loop Higgs-propagator corrections. The dot-dashed
line indicates the corresponding result where the box contributions have
been omitted, the dashed line shows the one-loop FD result, and 
the dotted line represents an improved Born approximation evaluated
within the renormalization-group-improved one-loop EPA.
\label{fig:higgsprod}
}
%\vspace{-2em}
\end{figure}

In \reffi{fig:higgsprod} the predictions for the production cross
sections $e^+e^- \to hZ, hA$ based on combining the complete
diagrammatic one-loop result in the MSSM~\cite{eehZA1l} with the
dominant two-loop Higgs-propagator corrections (evaluated with the
program \feh~\cite{feynhiggs})
are shown as a function of $\mh$ for $\sqrt{s} = 500$~GeV,
$\msusy = 1$~TeV, the slepton mass scale $\mslep = 300$~GeV, $\MA =
200$~GeV and $\Xt/\msusy = 2$, i.e.\ maximal mixing in the scalar top sector.
The result including the two-loop Higgs-propagator corrections is compared
with the one-loop result, and the effect of the one-loop box contributions
is shown separately. The FD result is furthermore 
compared with an improved Born approximation, where only corrections to
$\mh$ and $\aeff$ are taken into account, which are evaluated within the
renormalization-group-improved one-loop EPA with the program 
\subh\ (based on \citeres{mhiggsRGa,bse}).

%%%%%%%%%%%%%%%%%%%%% FIGURE %%%%%%%%%%%%%%%%%%%%%%%%%%%%%%%%%%%%%%%%%%
%%%%%%%%%%%%%%%%%%%%% FIGURE %%%%%%%%%%%%%%%%%%%%%%%%%%%%%%%%%%%%%%%%%%

As can be seen in the figure, the inclusion of the two-loop contributions
has a very large effect. The deviation between the full result and 
the improved Born approximation based on the renormalization-group-improved
one-loop EPA is also significant, exceeding 20\% for $\sigma_{\PA\Ph}$. 
The box contributions change the total cross section by 5--10\% at LC collider
energies, and in general give even larger corrections to the
differential cross sections. For the prospective experimental accuracies
at a future LC these corrections thus need to be included in the
theoretical predictions.

%%%%%%%%%%%%%%%%%%%%%%%%%%%%%%%%%%%%%%%%%%%%%%%%%%%%%%%%%%%%%%%%%%%%%
%%%%%%%%%%%%%%%%%%%%%%%%%%%%%%%%%%%%%%%%%%%%%%%%%%%%%%%%%%%%%%%%%%%%%

\end{document}